\documentclass[10pt, conference]{IEEEtran}
\usepackage[utf8]{inputenc}

\usepackage{cite}
\usepackage{multirow}
\usepackage{amsmath,amssymb,amsfonts}
\usepackage{algorithmic}
\usepackage{graphicx}
\usepackage{textcomp}
\usepackage{xcolor}
\usepackage{verbatim}
\usepackage{subcaption}
\usepackage{pgfplots, pgfplotstable}
\usepackage{tcolorbox}
\usepackage{makecell}
\usepackage{url}
\usepackage{caption}
\usepackage{subcaption}
\usepackage{graphicx}
\usetikzlibrary{positioning}
\usepackage{pgfplots}
\usepackage{array, tabularx, caption, boldline}
\begin{document}

\title{Towards Automatically Generating Release Notes using Extractive Summarization Technique}

\author{\IEEEauthorblockN{Sristy Sumana Nath}
\IEEEauthorblockA{Computer Science\\
University of Saskatchewan \\
Saskatoon, Saskatchewan \\
Email: sristy.sumana@usask.ca}
\and
\IEEEauthorblockN{Banani Roy}
\IEEEauthorblockA{Computer Science\\
University of Saskatchewan \\
Saskatoon, Saskatchewan \\
Email: banani.roy@usask.ca}}
\maketitle


\let\thefootnote\relax\footnotetext{DOI reference number: 10.18293/SEKE2021-119}
\begin{abstract}
Release notes are admitted as an essential document by practitioners.
They contain the summary of the source code changes for the software releases, such as issue fixes, added new features, and performance improvements. 
Manually producing release notes is a time-consuming and challenging task.
For that reason, sometimes developers neglect to write release notes. 
For example, we collect data from GitHub with over 1,900 releases, among them 37\% of the release notes are empty. We propose an automatic generate release notes approach based on the commit messages and merge pull-request (PR) titles to mitigate this problem.
We implement one of the popular extractive text summarization techniques, i.e., the TextRank algorithm. However, accurate keyword extraction is a vital issue in text processing. The keyword matching and topic extraction process of the TextRank algorithm ignores the semantic similarity among texts. To improve the keyword extraction method, we integrate the GloVe word embedding technique with TextRank. We develop a dataset with 1,213 release notes (after null filtering) and evaluate the generated release notes through the ROUGE metric and human evaluation. 
We also compare the performance of our technique with another popular extractive algorithm, latent semantic analysis (LSA).
Our evaluation results show that the improved TextRank method outperforms LSA.
\end{abstract}

\IEEEpeerreviewmaketitle
% \vspace{-3mm}

\section{Introduction}

Software release is a way to deliver the software package, with release notes, of the stable version to the end-users. 
Release notes are essential documentation in software development that contains a set of project activities, e.g., issue fixes, improvements and new features that have been implemented to a specific release \cite{arena1}. 
Different stakeholders (e.g., software development teams and external users) might benefit from release notes \cite{Semiautomatic, ARENA}. For example, the development team members (project manager, team lead, developer, tester) use them to learn what has changed in the source code to solve issues or integrate new features \cite{Semiautomatic}. Similarly, integrators, who are using a library in their code, use the library release notes to decide whether such a library should be upgraded to the latest release \cite{ARENA}. End-users and clients read the release notes to decide whether it would be worthwhile to upgrade to the latest software (e.g., application software or tool) version \cite{arnnode.js}. Besides, release notes serve as valuable resources to generate software documentation (e.g., software release report) submitted to the client. 
Moreover, practitioners use release notes in requirements engineering, software programming, software debugging and testing phase \cite{softwaredocumentation}.

\begin{figure}
    \centering
    \includegraphics[width=3.5in]{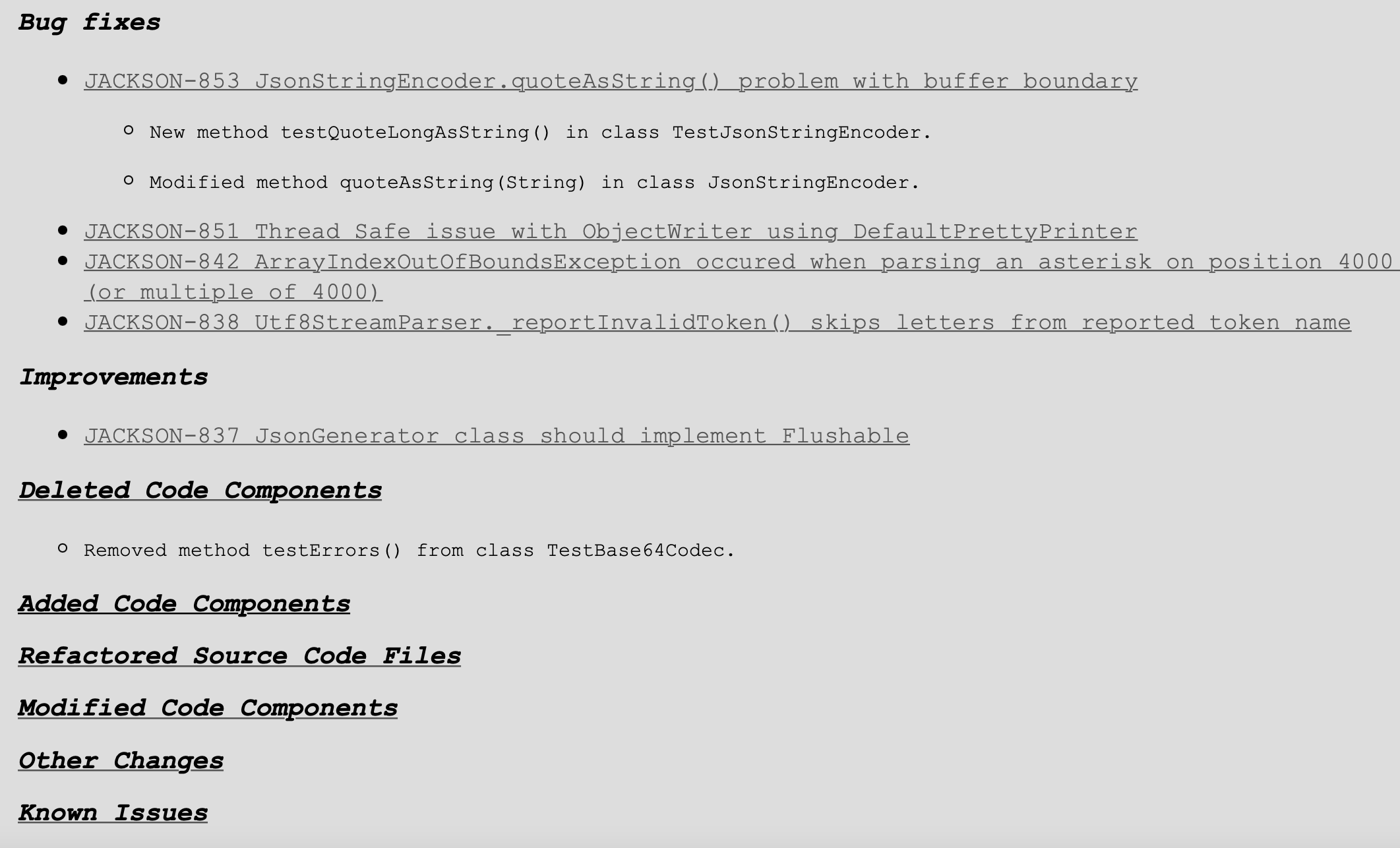}
    \caption{Generated Release Note by ARENA}
    \label{fig:arena}
\end{figure}

Generally, more than one developer is working on a single software project \cite{arnnode.js}. 
Generating release notes of a project by an individual is challenging as it is not feasible to know all fixed issues or the integrated new features of the project.
Therefore, primary release note producers collect information from contributors to produce the release notes manually \cite{rnempirical}. 
A few studies have proposed the generation of release notes - some of them are semi-automated \cite{Semiautomatic}, and some are automated techniques \cite{ARENA, arnnode.js} using manually-defined templates. The automated technique ARENA generates release notes by extracting source code change of Java projects and using some manually pre-defined templates based on the source code change type \cite{ARENA} (example in Figure \ref{fig:arena}). 
The primary limitation of this method is that it fails if the source code change or code statement is not matched with the pre-defined data, this approach cannot generate a sentence. Besides, this technique is only helpful for Java-based software. 
In an empirical study \cite{rnempirical}, 108 GitHub contributors and 206 IT professionals were surveyed for production and usage of release notes. The survey results reveal that none of the participants used ARENA for producing release notes.
Whereas the existing approach is not suitable to use; therefore, this is one of the motivations of our study.
On the other hand, the semi-automated technique \cite{Semiautomatic} filters good quality commit messages between two releases and appends them all to suggest producing release notes. Appending the commit messages of interrelated releases is not a feasible solution to generate release notes because it generates a lengthy release note. The audience wants to read precise release notes \cite{rnempirical}.
Some text management tools (e.g., issue trackers) are used to keep track of changes for producing and managing release notes, still need some considerable manual effort to create the release notes from these sources \cite{rnempirical}. 

% \vspace{-2mm}
\begin{table}[!ht]
        \caption{A Release Note in the Laravel Project}
    \label{tab:motivationexample}\centering
  \begin{tabularx}{8.5cm}{|X|}
    \hline
    \textbf{Release Note:}   \\
Add sanctum cookie endpoint to default cors paths \\
Modify the cache.php docblocks \\
Add stub handler \\
Closed \@auth correctly\\
        \hline
        \textbf{Commit messages \& merge PR titles:}\\
        Commit 1: Update CHANGELOG.md \\
        Commit 2: Merge branch `8.x' of github.com:laravel/laravel into 8.x \\
        Commit 3: Modify the cache.php docblocks \\
        Commit 4: add stub handler \\
         Commit 5: closed \@auth correctly \\
 Commit 6: add sanctum cookie endpoint to default cors paths \\
 Commit 7: add auth line \\
    \hline
  \end{tabularx}
\end{table}
% \vspace{-2mm}

Developing new tools could greatly help to improve release notes production and usage \cite{rnempirical}. As discussed above, there is a great lack of automatic release notes generation tools, which motivates us to develop an automated release notes generation technique. To work collaboratively, developers upload the source code in social coding platforms, e.g., GitHub, and can track the changes of the projects. A release is a collection of several commits, and commits can be regular commits and merge commits. Table \ref{tab:motivationexample} represents a release note of Laravel project\footnote{https://github.com/laravel/laravel/releases/tag/v8.4.2} with the related
commit messages and merge pull-request titles. Therefore, commit messages and merge pull-request titles are valuable artifacts to generate automated release notes.

We prepare a new dataset that contains the release information (e.g., release notes, release date) with their commit messages and merge pull-request titles of 13 projects from GitHub. After data filtering and pre-processing, our prepared dataset contains over 1,200 release data (detail described in Section \ref{dataset}). 
In our study, we regard the combination of commit messages and merge pull-request (PR) titles as an input source and release notes as the summary of the input sources.  We apply the TextRank algorithm; however, traditional TextRank algorithm ignores the semantic similarities. We integrate word embedding technique GloVe (Global Vectors for Word Representations) \cite{glove} to solve this problem and analyzed its effectiveness (described in Section \ref{resultandanalysis}).

To evaluate the automatically generated release notes, we also implemented the Latent Semantic Analysis (LSA) text summarization technique.
We evaluated automated release notes on the dataset using Recall-Oriented Understudy for Gisting Evaluation (ROUGE) \cite{ROUGE} and the F1-score obtained with ROUGE-1, ROUGE-2, and ROUGE-L is 31.74\%, 18.53\%, and 26.90\%, respectively. We also conduct a human evaluation to assess the quality of the generated release notes, which shows that our approach performs significantly better than the LSA and can generate more high-quality release notes.

In summary, our contributions are: 
\begin{itemize}
\item We develop a dataset with over 1,200 releases and their commit messages and merge pull-request titles from GitHub for release notes generation task. 
\item We implement the TextRank algorithm and integrate the GloVe word embedding technique to generate release notes for software release from their commit messages. 
\item We evaluate this approach on the dataset using the ROUGE metric and human evaluation. 
\end{itemize}

% \textbf{{ Structure of the paper:}}
We describe the research questions and usage scenarios in Section \ref{Motivation} and data collection process in Section \ref{dataset}. 
We elaborate our proposed approach in Section \ref{approach} and present the evaluation results in Section \ref{evaluation}. Section \ref{threat} and \ref{Study} discusses threats to validity and 
the related work. We conclude this paper with some future plans in Section \ref{Conclusion}.

\section{Study Design}\label{Motivation}
Our study  aims to answer the following two research questions:

\textbf{RQ1: What is the vital information that needs to include in the release notes?}
We investigate in the GitHub projects' repositories to understand the contents of the release note. We find that for the most cases
developers prepare a list of software changes of the current version from the last previous version. 
The list of software changes comes from commits and merge pull requests (Table \ref{tab:motivationexample}). For reason, we collect the commit messages and  merge pull requests titles as a \textit{input source} to generate automated release notes (detail describe in Section \ref{dataset}).

\textbf{RQ2: What is the efficient way to generate release notes?}
As per our previous discussion, it is clear that no appropriate tool exists for automated release notes generation in the software development practice. To produce concise release notes, we apply text summarization techniques. 
 We implemet the TextRank summarization algorithm by integrating the GloVe model. Using the GloVe model, we can resolve the semantic similarity issue of the traditional TextRank algorithm (detail describe in Section \ref{approach}). For evaluation, we have used the evaluation metric ROUGE to assess the text quality of the automated summary.
\subsubsection{{Usage Scenarios}}
The release notes may help developers and users to capture a summary of the latest release. Our approach can help developers to write the good quality of release notes. The usage scenarios are as follows:
\begin{itemize}
    \item It is challenging for developers to keep in mind all the changes for the upcoming release. Therefore, our approach can assist developers in producing release notes. 
    \item Our approach can help to replace the existing empty release notes in GitHub\footnote{https://github.com/django/django/releases}. 
\end{itemize}
\begin{figure}[!htbp]
    \centering
    \includegraphics[width=3in]{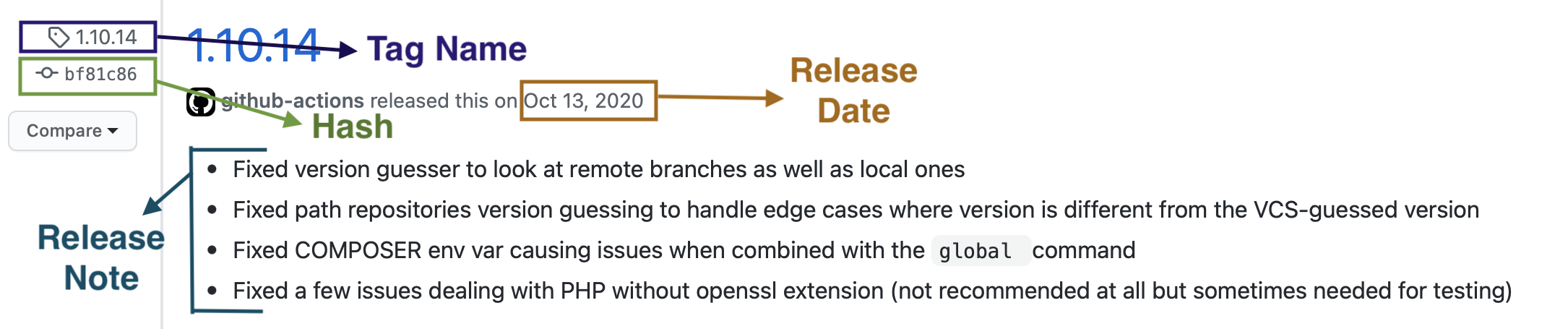}
    \caption{{Sample Release Note}}
    \label{fig:basicrelease}
\end{figure}

\subsubsection{{Problem Formulation}} 
Inspired from the example  shown in Table \ref{tab:motivationexample}, we observe the production of the release notes as a text summarization task with the combination of the commit messages and merge PR titles in the release as the \textit{article} and the release notes as the \textit{summary}. 
Therefore, the problem is formulated as follows: 

Given a source sequence $S_{seq}$ = $(x_1, x_2, ..., x_n)$, $n$th sentences in source sequence. 
An extractive summarizer aims to produce a summary $RN_{gen}$ by selecting $m$ sentences from $S_{seq}$, where $m$ is the length of existing release notes $RN_{ref}$. We use a scoring function $f()$ to generate a score for every sentence of $S_{seq}$ and then select top-ranked m sentences to produce an automated summary.

\section{Dataset} \label{dataset}
% ... \textcolor{red}{add flowchart}

\subsection{{{Data Preparation: }}}
In our study, we need to develop a dataset of release notes with commit messages and pull-request titles of the GitHub projects. To eliminate the trivial projects, we defined three criteria for project selection from GitHub projects: (i) the project is active (i.e., the repository of the project is being updated); (ii) the number of release notes in the project is more than 40; and (iii) the project have more than 8,000 stars.
Generally, GitHub projects are classified into six domains \cite{popularrepo}: (1) Application software, e.g., browsers, text editors; (2) System software, e.g., operating systems, (3) Web libraries and frameworks, e.g., Web API, (4) Non-web libraries and frameworks, e.g., android framework, 
(5) Software tools, like IDEs, and compilers;
(6) Documentation, like documentation, tutorials.

\begin{figure}
     \centering
     \begin{subfigure}[b]{0.4\textwidth}
         \centering
         \includegraphics[width=3in]{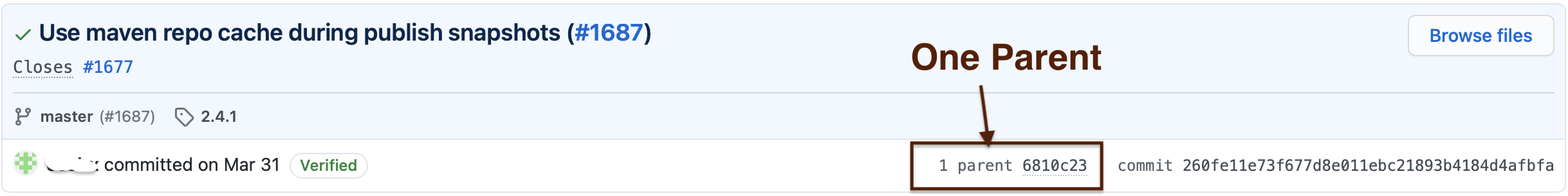}
         \caption{Regular Commit}
         \label{fig:regularcommit}
     \end{subfigure}
     \hfill
     \begin{subfigure}[b]{0.4\textwidth}
         \centering
         \includegraphics[width=3in]{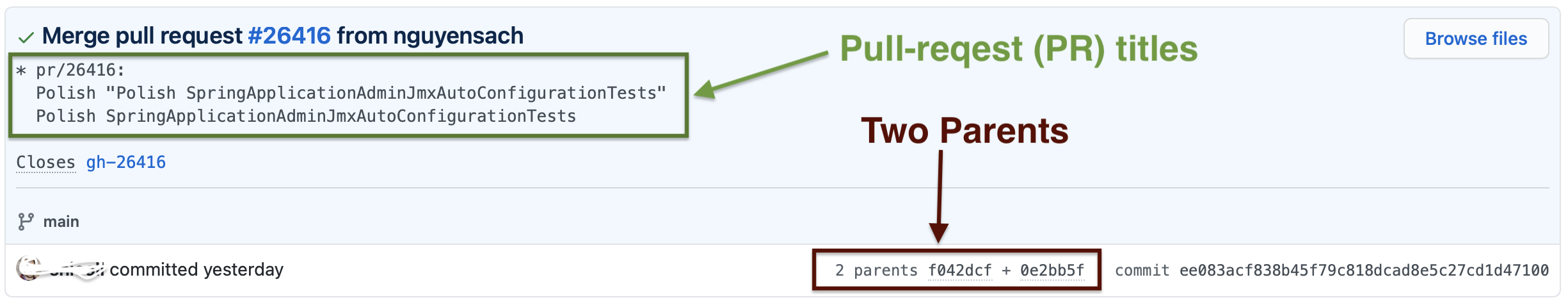}
         \caption{Merge PR Commit}
         \label{fig:mergedcommit}
     \end{subfigure}
     \hfill
        \caption{Commit}
        \label{fig:commit}
\end{figure}

\begin{figure}
    \centering
    \includegraphics[width=3in]{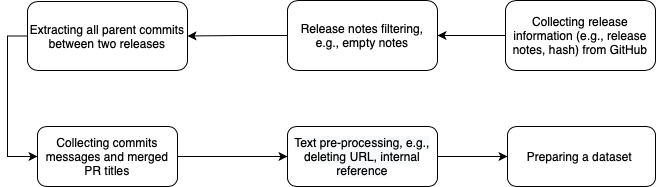}
    \caption{Data Collection Process}
    \label{fig:datacollection}
\end{figure}

We extract 1,924 release data (see in Figure \ref{fig:basicrelease}),
e.g., release note, hash, release data and tag name, of five domains except {Documentation} repositories of 13 projects 
% \cite{rxjava, cakephp,laravel,monolog,NewPipe, composer,matplotlib, keras, compose, dubbo, guava, redisson, exoPlayer}
(among them 6 Java, 3 Python and 4 PHP projects). We find 711 (37\%) release have empty release notes and develop a dataset with 1,213 release data. 
After that, we extracted the commit messages and merge pull-request titles of the releases. 
In GitHub, the new commit is linked with the previous commit using the parent, i.e., hash value (shown in Figure \ref{fig:commit}). We extract all parent hash values between two releases and then collect the commit messages and merge PR titles. Our study extracts the first sentences from the commit messages because the first sentences often are the summaries of the entire commit messages \cite{commitmsgNMT}. Similarly, merge PR commits have two parents (shown in Figure \ref{fig:mergedcommit}), one is previous commit and second is push PR commit. 
In our study, we collect all parents hash value between two releases and then extract the commit messages and merge PR titles using these hash values. 
% We use Python tool, PyDriller \cite{PyDriller}, for data mining.
Figure \ref{fig:datacollection} describes the process flow of data collection process. Data is uploaded in this link\footnote{https://github.com/sristysumana/SEKE2021Paper119}.


\subsection{{{Data Pre-processing: }}}
We conduct some text pre-processing techniques in the release notes and source sequences to filter out noise from the text. First, we eliminate the empty release notes. Then, we remove the HTML tags from the extracted release notes. Then, we split the text into sentences using NLTK (Natural Language Toolkit \cite{nltk}) and delete the a) url, b) reference number (e.g., “\#123”), c) signature, e,g, signed-off-by or co-authored-by, d) `@name', e) markdown headlines, e.g., \#\#\#Added, by identifying through regular expressions. Then, we filter out some trivial commit messages (for example, duplicate commit messages in same releases, `merge pull request/branch', `update .gitattributes' and so on).
\begin{figure*}[!htbp]
    \centering
    \includegraphics[width=6in]{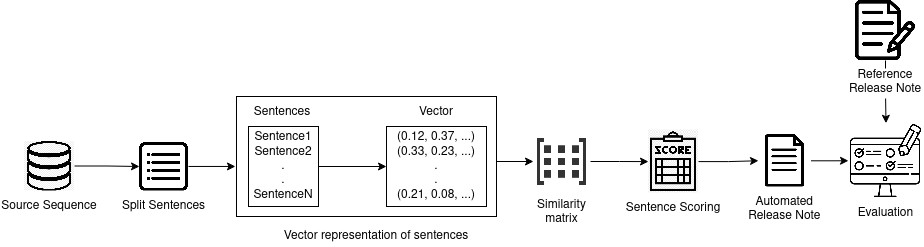}
    \caption{{Overview of the methodology}}
    \label{fig:textgeneration}
\end{figure*}

\section{Approach} \label{approach}
Automatic text summarization process produces a summary by reducing the text of a document. Text summarization approaches can be categorized into two types, extractive and abstractive summary. Extractive summary is created by extracting the phrases or sentences from the input sources and abstractive summarization process produces new words by understanding the main content of the input sources. We use extractive techniques in our study.
Usually, 
extractive summarization processes follow three steps to generate summary: (a) intermediate representation of the document, (b) scoring sentences and (c) strategies for selection of summary sentences.
The procedure of our approach is described in Figure \ref{fig:textgeneration}.
% In our study, we use TextRank text summarization algorithm to generate automated release notes for GitHub. 
\subsection{TextRank Algorithm}
TextRank is an unsupervised graph-based ranking algorithm, and it is widely used in keyword extraction and text summarization.
In general, graph-based ranking algorithms decide the importance of a vertex within a graph based on global information iteratively drawn from the entire graph. The core idea of the TextRank algorithm is to split the whole document into sentences. These sentences are used as nodes, and the weight between the nodes is used as edges. 
The weight between two nodes is calculated by similarity function and transform a similarity matrix among the sentences of the whole document. Then a text graph is formed, and the score (importance or rank) is computed of each vertex recursively until convergence.

Given, the source sequence, $D$, is set of $n$ sentences $\{S_1, S_2,...,S_n\}$ using NLTK package \cite{nltk}, i.e., $D = \{S_1, S_2,...,S_n\}$. By using each $S_i$ in $D$ as the node and the similarity between the nodes as the edge, a text graph $G = (D, E, W)$ can be constructed, where $E \subseteq SXS$ is nonempty finite set of edge between the nodes, $W$ is the weight set of edges, and $w_{ij}$ is the weight value of edges between node $S_i$ and $S_j$. The similarity function (Jaccard Similarity \cite{textrank}) for $S_i, S_j$ can be defined as:
% \vspace{-3mm}
\begin{equation}
    Sim(S_i,S_j) = \frac{|\{t_k|t_k \epsilon S_i \& t_k \epsilon S_j\}|}{log(|S_i|)+log(|S_j|}
\end{equation}
According to the $Sim(S_i,S_j)$ value, an similarity matrix $SM_{n\times n}$ is constructed between nodes.
The weight of elements on the diagonal of $SM_{n\times n}$ are all 1. Based on similarity matrix, the TextRank algorithm generates the importance or score of vertex $S_i$ using following equation,
\vspace{-2mm}
\begin{equation}\label{rankeq}
% \tiny
    S_c(S_i) = (1-d) + d*\sum_{S_j \epsilon In(S_i)} \frac{w_{ji}}{\sum\limits_{S_k \epsilon Out(S_j)}w_{jk}} Sc(S_j)
\end{equation}

Here, $S_c(S_i)$ is the ranking score of the node $S_i$, $d$ is the damping factor that represent the probability of the current node jumping to any other node, and at the same time can enable the weight to be transferred to the convergence stably. This co-efficient value can be set between 0 to 1, however $d$ usually set to 0.85 \cite{textrank, improvedTextRank}. $In(S_i)$ is the collection of all nodes pointing to node $S_i$, $Out(S_i)$ is the set of all nodes pointed by node $S_i$. The sum of the right side in the equation (\ref{rankeq}) indicates the contribution of each adjacent node to the node. The summed numerator $w_{ij}$ represents the degree of similarity between the two nodes $S_i$ and $S_j$ as well as the denominator is a weighted sum. $S_c(S_j)$ represents the weight value of node $S_j$ after the last iteration. After generating the score of all sentences, the TextRank algorithm generates a summary by selecting top-scored sentences.

\begin{table*}[!hbt]
\begin{center}
\caption{ROUGE scores}
\label{rouge}
    \begin{tabular}{  p{10em} |p{3em} | p{4em} | p{5em} |  p{3em} |p{4em} | p{5em} |p{3em} | p{4em} |p{5em} }
\hline
\multirow{2}{*}{\textbf{Approach}} & \multicolumn{3}{c|}{\textbf{ROUGE 1}} & \multicolumn{3}{c|}{\textbf{ROUGE 2}} & \multicolumn{3}{c}{\textbf{ROUGE-L}}\\ \cline{2-10} %\hline
& \textbf{Recall} & \textbf{Precision} & \textbf{F1-Score}& \textbf{Recall} & \textbf{Precision} & \textbf{F1-Score}& \textbf{Recall} & \textbf{Precision} & \textbf{F1-Score}\\   \hline
\textbf{LSA} & {28.65\%} & {30.28\%} & {29.44\%} & {14.89\%} & {14.35\%} & {14.61\%}& {23.54\%} & {25.45\%} & {24.45\%}\\   \hline
\textbf{TextRank+GloVe} & {\textbf{30.23}\%} & {\textbf{33.42\%}} & {\textbf{31.74\%}}& {\textbf{17.29\%}} & {\textbf{19.97\%}} & {\textbf{18.53\%}}& {\textbf{24.29}\%} & {\textbf{30.15\%}} & {\textbf{26.90\%}}\\   \hline
\end{tabular}
\end{center}
\end{table*}
\subsection{Text Vectorization}
Text Vectorization is the process of transforming text into numerical representation. 
Several ways are existing for the text vectorization, like, bag-of-words (BoWs), term frequency-inverse document frequency (TF-IDF), global vector for word representation (GloVe).
The traditional TextRank algorithm converts the document into BoWs vector representation, which is an unordered collection of word counts \cite{textrank} and the size of this vector is equal to the number of words in the vocabulary. The main issue of the BoW models is that if a sentence comes with the new words, then the vocabulary size would increase as well as the length of the vectors would increase. For that reason, the vector representation of the BoW will be a sparse matrix. Sparse representations are harder to model both for computational reasons and informational reasons as well as required a huge amount of training data.
To resolve this issue, the modified TextRank algorithm \cite{modifiedTextRank} integrates the TF-IDF algorithm. TF-IDF model reflects the importance of a word to a document in a collection or corpus. It can keep the relevant words score in the sentences and also considers the different lengths of the sentence in the document. TF measures the frequency of the word in each document in the corpus and IDF calculates the weight of rare words across all documents in the corpus. 
The formula of $TF-IDF(w_i)$ is as follows:
% \vspace{-1mm}
$$
TF(w_i) = \frac{f_{Sj}(w_i)}{f_{Sj}(w)} $$
$$
IDF(w_i) =\log \frac{N_s}{f_{S}(w_i)} $$
$$TF-IDF(w_i) = TF(w_i) \times IDF(w_i)
$$

where $f_{Sj}(w_i)$ represents the number of repetition of the word $w_i$ in the sentence $s_j$, $f_{Sj}(w)$ is the total number of words $w$ in the sentence $s_j$, $N_s$ represents the total number of sentences in the corpus; $f_{S}(w_i)$ represents the number of sentences containing the word $w_i$.

Still, capturing semantic similarity between the sentences is the issue of the TextRank algorithm. 
We observe that generating automated release notes needs to consider semantic similarity among the input sources (e.g., commit messages). Therefore, we integrate the GloVe word embedding model with the TextRank algorithm, which can improve keyword extraction performance by enhancing the semantic representations of documents. 
GloVe approach generates vector representation by calculating the  co-occurrence matrix of each word based on the frequency of word and co-occurrence counts.
It primarily gives information about the frequency of two words $Wi$ and $Wj$ appear together in the huge corpus.
To store this information, GloVe model generates co-occurrence matrix $X_w$, each entry of which corresponds to the number of times word $j$ occurs in the context of word $i$. As the consequence:
% \vspace{-1mm}
$$P_{{Wi}{Wj}} = P(Wj|Wi) = \frac{X_{{Wi}{Wj}}}{X_{Wi}}$$
is the probability that word with index $j$ occurs in the context of word $i$. This co-occurrence probability matrix gives a vector space with meaningful sub-structures.

\section{Evaluation Result \& Discussion}
\label{evaluation}
To compare the quality of generated release notes, we implement another text summarization technique named Latent Semantic Analysis (LSA) \cite{LSA}.
It is an unsupervised approach, not required to train the model. 
LSA extracts hidden semantic structures of words and sentences. 
An algebraic method, Singular Value Decomposition (SVD), is used to find out the interrelations between sentences and words.
% \subsection{Baselines}

\begin{table}[!hbt]
\begin{center}
\caption{Release Note Example I}
\label{example1}
    \begin{tabular}{ | p{30em}  |}
\hline

\textbf{Source Sequence:}
\begin{itemize}
\item made the parameter of {flowable,observable}.collect(collector) contravariant on t. 
\item 3.x: fix map() conditional chain causing npe 
    \item suppress undeliverable exception handling in tests.
\item fixed image link and added java examples for connectable observable operators.   
\item 3.x update conditional-and-boolean-operators.md
\item updating suppress undeliverable exception rule to have a named inner class instead of an anonymous inner class. 
\item edit dependency for gradle 

\end{itemize}
 \\ \hline

\textbf{Reference:}
\begin{itemize}
\item make the collector type of the collect operator contravariant in t
\item fix map() conditional chain causing npe
\item suppress undeliverable exception handling in tests
\end{itemize}
\\   \hline
\textbf{LSA:}
\begin{itemize}
\item suppress undeliverable exception handling in tests
 \item 3.x update conditional-and-boolean-operators.md
\item  updating suppress undeliverable exception to have a named inner class instead of an anonymous inner class \end{itemize}
\\   \hline
\textbf{TextRank+GloVe:}
\begin{itemize}
 \item 3.x: fix map() conditional chain causing npe
\item fixed image link and added java examples for connectable observable operators
\item suppress undeliverable exception handling in tests \end{itemize}
\\   \hline
\end{tabular}
\end{center}
\end{table}
% \vspace{-1mm}
\subsection{Evaluation Metric}
We evaluate the automated release notes with the ROUGE metric \cite{ROUGE}, which has been shown the quality of summarized text by high correlation with human assessments. Specifically, we use
ROUGE-N (N=1,2) and ROUGE-L, which are widely used to evaluate text summarization systems. 
The score of ROUGE-N is based on comparing n-grams in the generated and the original summary. The recall, precision and F1-score for ROUGE-N are calculated as follows:
$$
Recall_n = \frac{Count_{gram_n}(RN_{ref}, RN_{gen})}{Count_{gram_n}(RN_{ref})}
$$
$$
Precision_n = \frac{Count_{gram_n}(RN_{ref}, RN_{gen})}{Count_{gram_n}(RN_{gen})}
$$
$$
F1_n = 2\frac{Recall_n*Precision_n}{Recall_n+Precision_n}
$$
where, $Count_{gram_n}(RN_{ref}, RN_{gen})$ calculates the number of overlapping n-grams found in both the original and the generated text.
% $R_{rouge-n}$ measures the percentage of the n-grams in reference descriptions that an approach can generate, and $P_{rouge-n}$ presents the percentage of “correct” n-grams (i.e., n-grams appearing in reference descriptions) in generated descriptions. $F1_{rouge-n}$ is a summary measure that combines both precision and recall. 
The precision, recall and F1-score for ROUGE-L are similar with those for ROUGE-N, but instead of n-grams, they are calculated using the longest common subsequences between generated descriptions and reference descriptions \cite{ROUGE}. ROUGE is usually reported as a percentage value between 0 and 100. We obtained ROUGE scores using the `rouge-score' package \cite{rougescore} with Porter stemmer enabled.

\subsection{Experiment Settings}
We use \texttt{pandas}, Python library, for data manipulation and analysis. In text pre-processing phase, we use \texttt{nltk} library for tokenization and removing stopwords as well as write some regular expressions in the code to identify and delete the reference code, markdown, etc. from the text. In this study, we generate extractive summary for producing automated release notes. Therefore, we count the number of sentences (\textit{n}) of reference text and set \textit{n }is a parameter in the summarization technique. In order that, the sentence length of the reference and the generated release notes remain the same. For implementation, we use several library like, \texttt{gensim} \cite{genism}, \texttt{sumy} \cite{Sumy}, \texttt{sklearn} \cite{sklearn_api}, \texttt{networkx} \cite{NetworkX}.
\subsection{Analysis and Discussion} \label{resultandanalysis}
\subsubsection{{The Effectiveness of Our Approach}}To investigate the effectiveness of our approach (i.e., TextRank+GloVE), we evaluate the automated release notes with reference text on our dataset and the results are shown in Table \ref{rouge}. TextRank+GloVe model has a higher precision and F1 score than the LSA model for each ROUGE metric. The improvements compared with LSA in terms of the three F1 scores are 2.30, 3.91 and 2.44 points, respectively. These results indicate that compared with the two models, our approach can capture the key points of a release more precisely. 
The TextRank approach integrates the GloVe model, for that reason, this technique can consider semantic similarity between two sentences. On the other hand, sentence scoring and selection process of LSA depends on co-occurrence of words.
For example, LSA approach extract ``\textit{suppress undeliverable exception handling in tests}" and ``\textit{updating suppress undeliverable exception to have a named inner class instead of an anonymous inner class.}" in Table \ref{example1}. TextRank+GloVe can handle this situation by generating co-occurrence matrix of words based on the context.

\subsubsection{{The Effects of Main Components}}In this study, we generate release notes based on the TextRank approach by integrating  GloVe word embedding technique. We compare the automated release notes between TextRank+GloVe and TextRank+TF-IDF with the reference release notes (Table \ref{tab:example3}). TextRank+TF-IDF model extracts keywords based on the term frequency (TF) and inverse document frequency (IDF) and generate sentence score based on the high TF-IDF value. In release note example II (Table \ref{tab:example3}), TextRank+TF-IDF model identifies \textit{error, missing, components} as top keywords and selects top three scored sentences which have these keywords more than the other sentences. On the other hand, TextRank+GloVe extract better result than the TextRank+TF-IDF model. Because GloVe is a very powerful word vector learning technique which does not only rely on local context information of words, but also incorporates global word co-occurrence to obtain word vectors. For example, \textit{fix...replay} and \textit{fix...switchmap} are co-occurred in two sentences of source sequence in Table \ref{tab:example3}. Therefore, our model selects good sentences than the TextRank+TF-IDF. Table \ref{rouge2} shows the result by comparing the generated release notes by TextRank+TF-IDF and TextRank+GloVe model.
\begin{table}[!hbt]
\begin{center}
\caption{Release Note Example II}
\label{tab:example3}
    \begin{tabular}{ | p{30em}  |}
\hline

\textbf{Source Sequence:}
\begin{itemize}
    \item 3.x: add error handling section to observable.blockingfirst documentation 
\item 3.x: add missing coverage, update unused/inconsistent ops components
\item error occur when missing version tag from single.concatmaps 
\item add more tags those are missing previously
\item update 2.x maintenance date, include 3.0 wiki
\item update readme.md 
\item 3.x: fix switchmap not canceling properly during onnext-cancel races
\item 3.x: fix incorrect sync-fusion of switchmap and error management 
\item 3.x: fix replay not resetting when the connection is disposed
\item 3.x fix excess item retention in the other replay components 
% \item Fix typos in Backpressure.md 
\end{itemize}
 \\ \hline
% $\cdot$ test\\
\textbf{Reference:}
\begin{itemize}
  \item  fix switchmap not canceling properly during onnext-cancel races. 
  \item fix replay() not resetting when the connection is disposed. 
  \item add error handling section to observable.blockingfirst documentation.

\end{itemize}
\\   \hline
\textbf{TextRank + TF-IDF:}
\begin{itemize}
\item 3.x: add missing coverage, update unused/inconsistent ops components
\item error occur when missing version tag from single.concatmaps 
\item 3.x: fix incorrect sync-fusion of switchmap and error management 
% \item 3.x: Add `Error handling' section to Observable.blockingFirst documentation
\end{itemize}
\\   \hline
\textbf{TextRank + GloVe:}
\begin{itemize}
\item 3.x fix excess item retention in the other replay components 
\item 3.x: fix switchmap not canceling properly during onnext-cancel races
\item 3.x: add error handling section to observable.blockingfirst documentation

\end{itemize}
\\   \hline
\end{tabular}
\end{center}
\end{table}

\begin{table*}[!hbt]
\begin{center}
\caption{Effectiveness of the GloVe model}
\label{rouge2}
    \begin{tabular}{  p{10em} |p{3em} | p{4em} | p{5em} |  p{3em} |p{4em} | p{5em} |p{3em} | p{4em} |p{5em} }
\hline
\multirow{2}{*}{\textbf{Approach}} & \multicolumn{3}{c|}{\textbf{ROUGE 1}} & \multicolumn{3}{c|}{\textbf{ROUGE 2}} & \multicolumn{3}{c}{\textbf{ROUGE-L}}\\ \cline{2-10} %\hline
& \textbf{Recall} & \textbf{Precision} & \textbf{F1-Score}& \textbf{Recall} & \textbf{Precision} & \textbf{F1-Score}& \textbf{Recall} & \textbf{Precision} & \textbf{F1-Score}\\   \hline
\textbf{TextRank+TF-IDF} & {26.51\%} & {28.49\%} & {27.83\%} & {13.86\%} & {14.08\%} & {13.64\%}& {22.06\%} & {24.49\%} & {23.51\%}\\   \hline
\textbf{TextRank+GloVe} & {\textbf{30.23}\%} & {\textbf{33.42\%}} & {\textbf{31.74\%}}& {\textbf{17.29\%}} & {\textbf{19.97\%}} & {\textbf{18.53\%}}& {\textbf{24.29}\%} & {\textbf{30.15\%}} & {\textbf{26.90\%}}\\   \hline
\end{tabular}
\end{center}
\end{table*}

\subsubsection{{Where Does Our Approach Fail} }We carefully inspected the release notes where our approach does not obtain good F1 scores. We find that our approach usually performs badly because it presents the top ranking sentences and cannot generate new sentences. 
% if the reference description mainly contains information that cannot be found in the source sequence. 
The common case is that the source sequence contains several bug fix commit messages, however developers described \textit{`This release fixes few minor issues'} in the release notes to avoid a lengthy release notes. But our approach extracts a high ranked sentences from the source sequence. This issue is very common, hence it gets low ROUGE scores. Table \ref{tab:example2} represents a example where our approach fails. Additionally, the length of the generated text is a crucial matter for the poor score, because every sentence has a different word size. In our implementation, we only consider the total number of sentences, the ROUGE score would have been better if we had used the number of words to generate the summaries. However, in the extractive summary techniques, the word number cannot be used as a parameter. To reduce these limitations, we will implement sequence-to-sequence model to produce high-quality release notes in future and train the model.
% \vspace{-1mm}

\begin{table}[!hbt]
    \centering
    \caption{Release Note Example III }
    \label{tab:example2}
    \begin{tabular}{| p{26em}|}
    \hline
         \textbf{Source Sequence:}\\
         fixes failed to disable slave database and fixes unit test errors\\
fix oracle connection getschema() \\
fix shadow order\\
fix test cases and update encrypt strategy name\\
% fix MySQLNullBitmap\\
fix metric image path\\
fix integration test for select\_with\_case\_expression
         \\ \hline
         \textbf{Reference:}\\
         this release fixes few minor issues
         \\ \hline
         \textbf{TextRank+GolVe:}\\
         fixes failed to disable slave database and fixes unit test errors
         \\ \hline
        %  \textbf{Embedded-TextRank:}\\
         
        %  \\ \hline
    \end{tabular}
\end{table}

\subsection{Human Evaluation}
In this section, we invite 16 software engineers and get response from 13 participants to assess the quality of the generated release notes by LSA and TextRank algorithms. 
Automatic evaluation metric, ROUGE, calculates the textual similarity between the existing and the generated release notes, while the human study can evaluate the semantic similarity between them.
\subsubsection{{Procedure}} We also conduct a human evaluation to investigate our
approach’s effectiveness. We invited 13 human evaluators to assess the quality of the automated release notes. All of them are software developers with 3-8 years of experience and use GitHub.
We randomly select 10 release notes from the dataset and design an online survey. 
For each release, we show its original release notes followed by the two automated release notes generated by the TextRank+GloVe and LSA approaches to the evaluators. The two generated release notes are randomly ordered. Participants also have no idea about how these approaches work, so they cannot figure out which description is generated by which approach. 
Every evaluator provided a score from 1 to 7 to measure the semantic similarity between the generated release notes and the reference. The higher score means the existing release notes are closer to the automated summaries.
% \textcolor{red}{different type of extractive summarization, add figure}
\subsubsection{{Result}} 
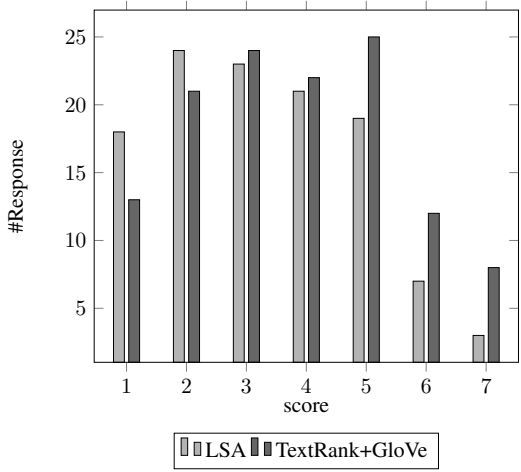
\begin{figure}
    \centering
    % \scriptsize
    \resizebox{.8\columnwidth}{!}{
  \begin{tikzpicture}
\begin{axis}[
ybar,
bar width=5pt,
	ylabel=\#Response,
	xlabel=score,
	enlargelimits=.09,
	xmin = 1,
	xmax = 7,xtick = data,
	legend style={at={(0.5,-0.2)},
	anchor=north,legend columns=-1}
]
\addplot [black,fill,fill opacity = .3] 
	coordinates { (1,18) (2,24)
		 (3,23) (4,21) (5,19) (6,7) (7,3)};
\addplot [black,fill,fill opacity = .6] 
	coordinates {(1,13) (2,21)
		 (3,24) (4,22) (5,25) (6,12) (7,8)};

\legend{ LSA, TextRank+GloVe}
\end{axis}
\end{tikzpicture}
}
  \caption{Score of Human Evaluation} 
  \label{fig:human} 
\end{figure}

Each release note obtains thirteen scores from
13 evaluators. We calculate the score and the
Figure \ref{fig:human} shows the final score of human evaluation. Each bar represents the obtained scores by the approaches in a specific score interval. For example, the right-most bar shows that 8 participants provide seven score to our approach. 
We notice that the generated release notes of TextRank+GloVe get more 4 to 7 score because it can extract most relevant sentences than the LSA.
% We can see that the release notes produced by TextRank are more close with the reference release notes compared to the LSA.
But we also notice that TextRank+GloVe generated release notes get the score between 1 and 2. The reason
may be that we do not train our model based on the existing release notes and hence sometimes may fail to generate high score of important sentences from source sequence.

% \begin{figure}
%     \centering
%     % \includegraphics[width=1\linewidth]{Conference-LaTeX-template_7-9-18/images/New Method Count.png} 
%   \begin{tikzpicture}
% \begin{axis}[
% ybar,
% 	x tick label style={
% 		/pgf/number format/1000 sep=},
% 	ylabel=\#Response,
% 	xlabel=score,
% 	enlargelimits=0.03,
% 	legend style={at={(0.5,-0.2)},
% 	anchor=north,legend columns=-1}
% ]
% \addplot [black,fill,fill opacity = .3] 
% 	coordinates { (1,14) (2,16)
% 		 (3,18) (4,15) (5,11) (6,7) (7,3)};
% \addplot [black,fill,fill opacity = .6] 
% 	coordinates {(1,13) (2,16)
% 		 (3,16) (4,20) (5,18) (6,10) (7,5)};

% \legend{ LSA, TextRank+GloVe}
% \end{axis}
% \end{tikzpicture}

%   \caption{Score of Human Evaluation.} 
%   \label{fig:human} 
% \end{figure}

\section{Threat to Validity}\label{threat}
One threat to validity is that our dataset is built from Java, Python and PHP projects in GitHub repositories. Therefore, it may not represent all programming languages. Besides, our approach takes commit messages and merge pull-request titles as input; hence can also be applied to projects of other programming languages.

Another threat to validity is that the non-summary information, such as acknowledgment of contributors, installation command for the new release, in release notes may affect the effectiveness of our approach. Release notes are free-form text, and we cannot guarantee their quality and content. We try to mitigate this threat by using a set of heuristic rules and manual analysis to filter out non-summary information when pre-processing. However, it is hard to process the patterns of all non-summary information. For future work, we will focus on data pre-processing for further improvements.

Another threat is we cannot compare the generated release notes from ARENA and our approach because ARENA extracts all issues from JIRA to generate release notes for Java projects and generate lengthy release notes. Moreover, from survey study, no participants have adopted ARENA for producing release notes \cite{rnempirical}.

There is also a threat related to our human evaluation.
We cannot guarantee that each score assigned to every release note is fair. To mitigate this threat, we invite the professional developers to our survey who have experience in GitHub and producing release notes.
% each sampled release note is evaluated by 13 evaluators, and we use the average score of the 13 evaluators as the final score.

\section{Related Study}\label{Study}
Several empirical studies focused on the usage of release notes.
Abebe et al. \cite{releasenote} identify nine important factors (e.g., issue priority and type) in explaining the likehood of an issue being included in release notes.
Tingting et al. \cite{rnempirical} analyzed 32,425 release notes and classified the content into eight categories, e.g., 
issues fixed, new features, and internal changes. Among them, \textit{issues fixed} (79.3\% of the release notes) and \textit{new features} (55.1\% of the release notes) are the most documented categories. In GitHub, developers push commits or send pull requests in a separate branch to resolve the issues. 
For that reason, we propose an automated release notes generation technique based on the commit messages and merge PR titles. 

Some other studies aimed to generate automated release notes. Moreno et al. \cite{ARENA} propose ARENA tool to produce release notes for Java projects. ARENA extract the source code changes from GitHub and collect issues from JIRA using issue-commit linker. Then it prepare a list of issues and generate change description using predefined text templates. The example is shown in Figure \ref{fig:arena}. 
Similarly, Ali et al. applied the same technique to generate automated RNs for Node.js projects \cite{arnnode.js}. 
Klepper et al. \cite{Semiautomatic} proposed a semi-automatic approach by collecting information from the build server, issue tracker and version control system. 
Then, release manager can edit the this list before publishing. However, we implemented full-automatic approach to produce release notes and pre-defined text templates is not required in our study.

% \subsection{Documenting Other Software Artifacts: }
For automated software documentation, researchers studies different types of software artifacts, such as commit messages generation \cite{changescribe, commitmsgNMT} and automated pull-requests description \cite{prdescription}. % For example, code diff are used as a source sequence to generate commit messages using manually-defined templates \cite{changescribe} and  neural machine translator \cite{commitmsgNMT}. 
For example, Liu et al. \cite{prdescription}
propose an automatic approach to generate PR descriptions based on the commit messages and the source code comments. On the other hand, release combines plenty of commits and pull requests. 
% Commit messages and pull-request titles are usually helpful for developers, while release notes are generally prepared for software development team members and end-users. 
Our work aims to generate release notes from commit messages and pull-request titles using an extractive method. 

\section{Conclusion \& Future Work }\label{Conclusion}
This paper aims to generate release notes for software releases automatically. 
% Generally, a release note contains the main resolved issues of the specific release. Table \ref{tab:motivationexample} represents a sample of release notes with their commits, and the documented information of release notes is very close to the commits messages and merge PR titles. 
In our work, we apply the extractive text summarization technique, i.e., TextRank, to produce automated release notes by selecting top-ranking sentences from their commit messages and merge PR titles.
The main novelty of this work is that we do not use any pre-defined template like ARENA \cite{ARENA}. Our approach is language-independent; oppositely, ARENA is used for Java projects.
Moreover, we integrate GloVe to convert the text into embeddings, which helps to extract keywords than the traditional approach. We evaluate the generated release notes using the ROUGE metric and conduct a human evaluation to check the effectiveness of the automated release note generation technique.

In future, we plan to apply a sequence-to-sequence model to improve the quality of release notes. We also plan to improve our approach by involving additional related software artifacts as input. For example, by taking git diff, relevant content from bug reports as input and text summarization models may be able to infer the implementation details and the motivation of a software release. Moreover, we plan to improve the structure of release notes. We will categorize, e.g., bug fixes, improvements, new features, the content of release notes for better representation to release note users.

\section*{Acknowledgment}
This research is supported by the Natural Sciences and Engineering Research
Council of Canada (NSERC), and by two Canada First Research Excellence
Fund (CFREF) grants coordinated by the Global Institute for Food Security
(GIFS) and the Global Institute for Water Security (GIWS).
% \nocite{*}
\bibliographystyle{plain}
\bibliography{main}
\end{document}